# Neural Optimization with Adaptive Heuristics for Intelligent Marketing System

Changshuai Wei[*]
LinkedIn Corporation
Seattle, USA
chawei@linkedin.com

Benjamin Zelditch
LinkedIn Corporation
New York, USA
bzelditch@linkedin.com

Joyce Chen
LinkedIn Corporation
Sunnyvale, USA
joychen@linkedin.com

Andre Assuncao Silva T Ribeiro
LinkedIn Corporation
Sunnyvale, USA
aribeiro@linkedin.com

Jingyi Kenneth Tay
LinkedIn Corporation
Sunnyvale, USA
ktay@linkedin.com

Borja Ocejo Elizondo
LinkedIn Corporation
Sunnyvale, USA
bocejo@linkedin.com

Sathiya Keerthi Selvaraj
LinkedIn Corporation
Sunnyvale, USA
keselvaraj@linkedin.com

Aman Gupta
LinkedIn Corporation
Sunnyvale, USA
amagupta@linkedin.com

Licurgo Benemann De Almeida
LinkedIn Corporation
New York, USA
lalmeida@linkedin.com

## ABSTRACT

Computational marketing has become increasingly important in today's digital world, facing challenges such as massive heterogeneous data, multi-channel customer journeys, and limited marketing budgets. In this paper, we propose a general framework for marketing AI systems, the Neural Optimization with Adaptive Heuristics (NOAH) framework. NOAH is the first general framework for marketing optimization that considers both to-business (2B) and to-consumer (2C) products, as well as both owned and paid channels. We describe key modules of the NOAH framework, including prediction, optimization, and adaptive heuristics, providing examples for bidding and content optimization. We then detail the successful application of NOAH to LinkedIn's email marketing system, showcasing significant wins over the legacy ranking system. Additionally, we share details and insights that are broadly useful, particularly on: (i) addressing delayed feedback with lifetime value, (ii) performing large-scale linear programming with randomization, (iii) improving retrieval with audience expansion, (iv) reducing signal dilution in targeting tests, and (v) handling zero-inflated heavy-tail metrics in statistical testing.

## CCS CONCEPTS

• **Computing methodologies** → **Machine learning**; • **Applied computing** → **Marketing**; • **Mathematics of computing** → *Probability and statistics*; *Mathematical optimization*.

[*]Corresponding author.

## KEYWORDS

Computational Marketing, Linear Programming, Lifetime Value, Audience Expansion

## 1 INTRODUCTION

Compared with traditional media marketing, online digital marketing has become increasingly important for advertisers and marketers. With online digital marketing, marketers can reach customers more directly with personalized marketing messages. Many of the current major internet companies build advertising platforms to better connect advertisers/marketers to customers, typically through auction or bidding systems. From a marketer's perspective, there are paid social channels (e.g., Facebook) and paid search channels (e.g., Google). Besides these paid channels, there are also owned channels like email, where marketers can connect with customers directly. There is much literature on building machine learning (ML) systems from ads platform's perspective; however, few papers have discussed ML and optimization systematically from the marketer's perspective. In fact, there are many unique challenges when building an intelligent marketing system from the marketer's perspective:

(1) **The data can be massive and heterogeneous.** Since there are multiple marketing channels, the granularity of data from different channels are highly heterogeneous. For owned channels like email, marketers can collect member-level user feedback, whereas for paid channels like Google Search, marketers can only collect aggregate-level information, e.g., at the keyword or campaign level.

(2) **The objective can be challenging to define.** The marketing life-cycle can be quite different across products. For many to-consumer (2C) products, the interaction is typically online and self-served, so the life-cycle from receiving the marketing message to conversion can be just a few days. For to-business (2B) products, the life-cycle can span weeks to months, including activities such as lead generation and sales involvement. As a result, these ML models face issues associated with delayed feedback.

(3) **The action space is complex.** There are different types of actions to optimize the marketing outcome. These include (but are not limited to) at what time and frequency the marketing message should be sent, what content (e.g., picture, language) to use for the message, which audience to target, how much to bid for the marketing campaign(s) on various ads platforms. Unifying these actions under a general optimization framework would help build a more efficient marketing system.

(4) **There are various types of constraints.** Marketing typically comes with cost. Some costs are explicit (e.g., billing from paid channels), while others are implicit (e.g., complaints or unsubscriptions from users on owned channels). Besides these cost constraints, there are also other operational constraints within marketing systems, e.g., capping on message frequency, minimum volume for various type of products. An intelligent marketing system should be able to determine optimal actions while respecting these constraints.

To address these challenges, we propose *Neural Optimization with Adaptive Heuristics (NOAH)*, a general framework for intelligent marketing systems. Before we introduce the framework, we briefly review related ML and optimization works in the marketing industry and for general recommender systems.

## 1.1 Related Work

*1.1.1 ML in Marketing Industry.* The use of ML in marketing has been fertile over the last decades, given the wealth of data naturally produced by marketing and the capacity of ML tools/methods to transform data into information that can be used for marketing optimization and decision making. For instance, ML solutions have been proposed and/or applied to multiple domains, for instance: customer segmentation in industries such as retail [13, 31], hospitality [1, 41] and banking [32, 42]; customer life time value (LTV) [18] in industries ranging from gaming [11], to e-commerce [21], to subscription services [27]; marketing attribution, in both multi-touch attribution [37] and marketing mix modelling [22]; the optimization of different marketing channels, such as email [30, 38], paid search bidding [15, 17], display advertising [35, 39], among others. For a comprehensive review of the use of ML in marketing, we recommend the work presented in [25, 28].

*1.1.2 Optimization in Recommender Systems.* A commonly used optimization approach for Recommender systems (RecSys) is the ranking heuristic, where items are ranked by their relevance scores and top-ranked items are shown/sent to users. To balance different (and sometimes conflicting) objectives, the relevance score is often a linear combination of multiple predicted objective values. While a large amount of research effort has gone into applying ML and AI to predicting the individual objective values, less work has been done on how to choose the weights for an optimized linear combination. In recent years, Linear Programming (LP) has been emerging in RecSys for its capability of trading off multiple objectives.

LP has been studied closely since the 1940s, and several commercial solvers such as Gurobi [19] and GLPK [29] can give high-quality solutions for small-moderate instances in reasonable time frames. However, they cannot scale to industrial-scale problems with billions or trillions of variables due to memory or computational constraints. One approach to handle large-scale LPs is to devise algorithms where certain key quantities can be computed in a distributed manner. The alternating direction method of multipliers (ADMM) [7] is one such approach; [43] details how ADMM can be implemented successfully for industrial-scale problems. Another approach is the primal-dual hybrid gradient (PDHG) method [9] for a class of problems in convex optimization. [2] applies the PDHG method to a saddle-point formulation of LP and adds enhancements such as diagonal preconditioning, presolving, adaptive step sizes and adaptive restarting to make the algorithm more efficient in practice.

## 1.2 Our Approach

NOAH is a general framework for intelligent marketing systems which consists of three major modules: 1) prediction module, 2) optimization module and 3) adaptive heuristics module. The prediction module provides predictions of business metrics given possible marketing actions. The optimization module chooses marketing actions that optimize the relevant business metrics. The adaptive heuristics module connects various components of the system to form a better feedback loop. We first go over more details of the NOAH framework in Section 2, including the three NOAH modules and some illustrative high-level examples. We then walk through a large scale application to LinkedIn's marketing engine in Section 3. We also share the real A/B test win results with some learnings in Section 4 and conclude in Section 5. We would like to highlight two major contributions of the paper:

(1) As far as we know, this is the first general framework for marketing optimization that takes into account both 2B and 2C products, as well as both owned and paid channels.
(2) This is a large-scale application to LinkedIn marketing that led to a major business win, and we report details of our learnings and experiences in this paper.

These contributions imply that our approach offers great scope for improving marketing optimization in general and hence is broadly useful.

## 2 NOAH FRAMEWORK

In this section, we will first give a brief overview of LinkedIn's marketing system. We then introduce the details of different modules and components of the NOAH framework, followed by application guidelines and high-level examples for bidding and content optimization.

## 2.1 Marketing Ecosystem at LinkedIn

LinkedIn is the world's largest professional network and has products that serve both 2C and 2B needs. As a result, the marketing system (Figure 1) needs to promote not only 2C products like Premium but also various 2B products in Talents, Learning, Ads and Sales lines of business. The marketing system is a heterogeneous system that involves both a first-party system and integration with third-party tools. Marketers interact with the marketing system to decide the optimal marketing actions for these 2B and 2C products, including whom the marketing message should be sent to, the best timing and frequencies for the messages to be sent, what bidding price or target the keywords or campaigns should be set at (on paid channel). Once the actions are decided from the Marketing Engine, the marketing messages are delivered via various owned or paid channels.

Paid channels include paid search, paid social and other display ads on third-party platforms, where LinkedIn (as an advertiser) participates in ad bidding through various auction mechanisms. Meanwhile, with owned channels, LinkedIn can promote its products on LinkedIn itself or use the email channel to have direct connections with customers. When a user clicks on a marketing message from one of these channels, they are directed to an optimized landing page where more detailed product information is shown. For 2C products, the landing page is typically on LinkedIn's main website and the purchase can be typically completed in a self-serve manner. On the other hand, for 2B products the landing page is typically Linkedin's 2B micro-site (e.g., https://business.linkedin.com/) where an inquiry form can be generated for LinkedIn sales personnel to follow up.

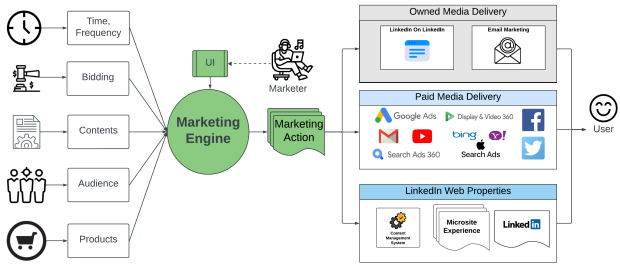

Figure 1: Overview of LinkedIn's Marketing System.

## 2.2 NOAH Modules

In order to have a general framework that can handle different types of entities and possible actions in owned and paid channels, we first abstract out two definitions: *marketing unit* and a *marketing action*. A *marketing unit* is an entity that a *marketing action* can be applied on. For example, in email marketing where we can directly connect with members and collect engagement data, the marketing unit is a member and a marketing action (to be optimized) can be whether to send a particular email to the member. In paid search channels, assuming marketers can get keyword-level data, the marketing unit can be a keyword, and a marketing action can be the bid amount for the keyword.

Figure 2 shows the overall diagram of the NOAH system. The core engine (inside the large dashed green box) consists of three key modules: (i) prediction, (ii) optimization, and (iii) adaptive heuristics. We first introduce the components outside the core engine so that the context, inputs and outputs around the core engine are clear.

NOAH is a human-in-the-loop system. Marketers with expert knowledge define appropriate marketing units and marketing actions in the relevant databases. A candidate retrieval process retrieves the appropriate set of marketing units and actions from the databases and feeds them to the core engine. Marketers also provide input parameters via an UI that can guide the core engine's behavior. User engagement feedback and other contextual information are also collected as input to the core engine. The core engine processes these inputs through the three key modules to generate optimized marketing actions.

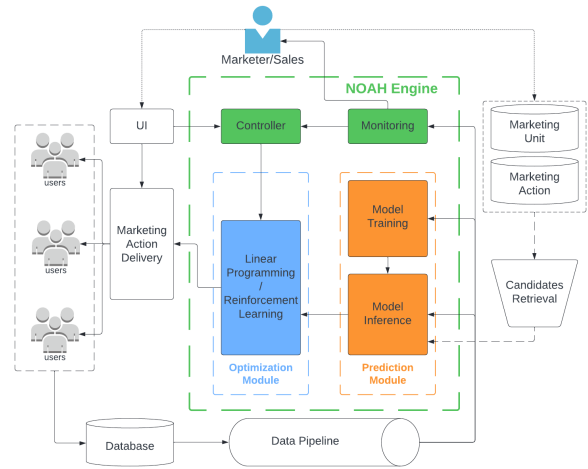

Figure 2: NOAH Framework.

*2.2.1 Prediction Module.* In prediction module, we want to predict performance metrics for each marketing unit given some marketing action. Denoting the feature vector for marketing unit $u$ as $x_u$, feature vector for marketing action $a$ as $x_a$ and a metric as $y$, the prediction module builds a functional mapping:

$$y = f(x_u, x_a, x_c),$$

where $x_c$ represents a feature vector for contextual information other than marketing unit and action (e.g., location, device). The metrics are typically marketing objectives and constraints related to user feedback. For example, we can build a functional mapping for four metrics: click through rate (CTR), conversion rate, profit per conversion and cost per click (CPC). With these metrics, we can use the chain rule to connect these conditional probabilities to estimate the profit per impression and cost per impression, which can be further used in the optimization module. The functional mapping $f(\cdot)$ can be learnt from user feedback data. Depending on the amount of data available and the maturity of the system, we can use different supervised learning methods to estimate $f(\cdot)$. If the amount of data is small and the project is at the initial stage, a simple regression model is often enough. If more data is available and the project is at the initial stage, an off-the-shelf gradient boosted tree [16] or XGBoost model [12] can give decent performance. If data is abundant and the project is in the iteration phase, we have found that deep neural networks give the best performance due to their flexible architecture. For example, transformer architectures can be used to generate embeddings for historical marketing interaction with users. If multiple metrics are being modeled, multi-task learning architectures like multi-gate mixture-of-experts (MMOE [24]) or entire space multi-task model (ESMM [26]) can be employed to exploit the relationship among multiple metrics during training.

One key characteristic of the prediction module is the inclusion of the marketing action. If the metric prediction is conditioned on the marketing action, the prediction module can then generate metric estimates for all possible marketing actions, from which the optimization module can choose the best action. However, historical data may not contain all possible marketing actions for each unit, which may generate bias in inference[10]. To reduce the bias, various causal ML methods can be used so that the $f(\cdot)$ learnt better reflects the causal relationship [14, 20].

*2.2.2 Optimization Module.* With the functional mapping of marketing actions and performance metrics, we then need to choose the best action for a given marketing unit in the optimization module. The simplest, widely adopted approach is a ranking approach: choose the action with the largest $y$ value, i.e., the largest estimated metric. While it can often work, this ranking approach ignores many limitations or constraints that can lead to sub-optimal and sometimes unacceptable performance.

To address various constraints across marketing units, we can frame the problem as a constrained optimization problem. For each metric $k = 0, 1, \ldots, K$, let $y^k = f_k(x_u, x_{a_u}, x_c)$ denote the estimated value of metric $k$ for marketing unit $u$ under action $a_u$. Without loss of generality, let $y^0$ be the primary metric to be optimized, and let the remaining $y^k$ ($k \geq 1$) be metrics that are subject to various limitations or business constraints with upper bound constraints $C_k$. Assuming no interaction effects across marketing units, we can formulate the optimization problem:

$$\max_{a_u} \sum_u f_0(x_u, x_{a_u}, x_c), \tag{1}$$

$$\text{s.t.} \sum_u f_k(x_u, x_{a_u}, x_c) \leq C_k, \quad \forall 1 \leq k \leq K, \tag{2}$$

The above optimization problem can be relaxed and solved with efficient linear programming (LP) algorithm in many cases; we detail some examples in Section 2.3 and Section 3.3. When latency is a strong requirement, some compromises on accuracy need to be made, for example, to use yesterday's dual solution for today's primal calculation. Fortunately, for marketing AI use cases, latency is not a major problem as the systems are mostly offline. For example, in email marketing, marketers control the time and frequency and typically limit the frequency to avoid spamming users. In paid media, marketers only need to change the non-impression-level bid, e.g., cost-per-click (CPC) bid as the impression-level response is taken care of by the ads platform. For these reasons, LP is well suited for marketing AI use cases.

Besides the marketing cost and business constraints on the "space" dimensions, there are implicit constraints on the "time" dimension, particularly for each marketing unit. The time window to attract and convert a potential user is typically limited. Moreover, a marketing action taken at a given time point can impact the status and best marketing action to take at the next time point. We can leverage Reinforcement Learning (RL) to address optimization on "time" dimension. Contextual bandits can be used to balance the exploration and exploitation. Q-Learning can be used to help find an optimal sequence of marketing actions to maximize value for a given marketing unit.

To summarize, the optimization module addresses the resource limitation problem. As a general principle, we recommend LP when the resource limitation across marketing units is strict and there are less frequent marketing interactions, and recommend RL when there are more frequent marketing interactions and less constraints across marketing units (or these constraints can be taken care of by other parts of the system). We note that these two techniques are not mutually exclusive and can be combined. For example, we can focus on "space" constraints and use an LP formulation, while simplifying the time horizons resolution to an LTV prediction. We can also focus on "time" horizons and use an Q-learning formulation, while transforming the cost constraints implicitly to reward shaping. We argue that the former combination are often preferred in marketing, since marketing use cases are typically less frequent interaction with strict cost constraints.

*2.2.3 Adaptive Heuristics Module.* Adaptive heuristics (AH) are lightweight components that help the whole system run stably and coherently, each of which may have some prediction and optimization functionality. We here briefly discuss two main classes of AHs and provide more details in Appendix B.

One important class of AH is *Feedback Loop Controllers* for business constraints (green boxes as shown in Figure 2). Since the market is dynamic and the marketing AI system can have errors (e.g., prediction error), the observed cost $C'$ may be greater (violation) or much less (under-utilization) than the budget $C$. To minimize under- or over-utilization, we can continuously monitor the performance and build a lightweight controller, $C^* = g(C, C')$, to update $C^*$ as the input to the optimization module instead of the original $C$.

Another important class of AH is *Hierarchical Optimization*, so that the aggregate-level constraints propagate to lower level. For example, a macro-level optimization with a forecasting model and an LP can be built to allocate marketing budget to different marketing channels and products.

## 2.3 Guidelines and Examples

NOAH is a general framework for marketing AI systems. Ideally, we would like to perform joint optimization for multiple types of marketing actions, e.g., product, timing, frequencies and content. However, this is often infeasible due to reasons such as infrastructure constraints, operational convenience, or limitations in the ML models or optimization solvers. As a workaround, we can use the same NOAH framework and perform optimization in a cascading/sequential manner for different types of marketing actions. For example, we can first decide the product for each marketing unit, then optimize the timing and frequencies, and finally choose the best content. Another benefit of this muti-stage design is that it allows for a flexible combination of different optimization techniques for different actions. Here, we give examples for two different marketing actions: bidding and content optimization.

*2.3.1 Bidding Optimization Example.* For bidding, advertisers can receive hourly or daily ads winning performance metrics $x_p$, for example, average position for each keyword, or impression winning percentage for each campaign. One can build a series of models to predict: number of clicks $y^{\text{clicks}} = f_{\text{clicks}}(x_u, x_p)$, cost per click $y^{\text{cpc}} = f_{\text{cpc}}(x_u, x_p)$, conversion rate $y^{\text{cvr}} = f_{\text{cvr}}(x_u)$ and revenue per conversion $y^{\text{rpc}} = f_{\text{rpc}}(x_u)$. Note that we can assume $y^{\text{cvr}}$ and $y^{\text{rpc}}$ do not depend on ads winning performance as they happen after users click the ad and land on the landing page. Typically, advertisers want to maximize revenue while maintaining certain level of return-on-ad-spends (ROAS). The corresponding constrained optimization formulation could be:

$$\max_{p(u)} \sum_u y^{\text{rev}}_{u,p(u)},$$

$$\text{s.t.} \frac{\sum_u y^{\text{rev}}_{u,p(u)}}{\sum_u y^{\text{cost}}_{u,p(u)}} \geq C_{ROAS},$$

where $p$ represents the index for categorized winning performance, and

$$y^{\text{rev}}_{u,p} = f_{\text{clicks}}(x_u, x_p) f_{\text{cvr}}(x_u) f_{\text{rpc}}(x_u),$$
$$y^{\text{cost}}_{u,p} = f_{\text{clicks}}(x_u, x_p) f_{\text{cpc}}(x_u, x_p).$$

The constrained optimization problem can be solved with the efficient algorithm in Appendix A.1 if the scale is large. Once the desired optimal $x_p^*$ is obtained, we can calculate the optimal CPC bid as $a_u^* = f_{\text{cpc}}(x_u, x_p^*)$, which can be directly used in first price auction platforms, or as the expected-cost-per-click (ECPC) bid in second price auction platforms.

*2.3.2 Content Optimization Example.* Content optimization can happen at different stages across different marketing channels. Without loss of generality, assume there is a slot on the landing page where we can show different creatives $a$ to different users $u$. We can build an online high-dimensional contextual bandit model based on following regression model:

$$\rho^{-1}(y) = x^T w + \epsilon, \qquad \epsilon \sim N(0, \beta^2),$$

where $x = (x_u, x_a, x_{u,a})$ is feature vector that includes features for user $x_u$, features for creative $x_a$, and features for the user-creative interaction $x_{u,a}$, $\rho(\cdot)$ is a link function, and $y$ is observed reward. Online updates for the posterior distribution of $w_i \sim N(\mu_{w_i}, \sigma^2_{w_i})$ can be computed efficiently using factor graph and expectation propagation (see algorithm details in Appendix A.2). Thompson sampling can then be employed to obtain predicted rewards for different creatives, and show users the creative with the largest reward on the landing pages.

# 3 EMAIL MARKETING APPLICATION

In this section, we report the details of a large-scale application of the NOAH framework to LinkedIn's email marketing engine, including important innovations and learnings.

## 3.1 Email System

We can think of the email marketing system as a two-stage recommender system: candidates retrieval followed by email optimization (Figure 6 in Appendix). In the candidate retrieval stage, marketers create email marketing campaigns for different products and define the audience list for each campaign. In the email optimization stage, ML models are used to find the best marketing emails for a selected group of members, so as to optimize metrics while minimizing negative member experiences like spamming. The whole system runs in batches, with emails sent out at regular time intervals (e.g., weekly). The system seeks to maximize value for LinkedIn while respecting multiple business constraints. Some constraints are global, e.g., the total number of user unsubscriptions should be less than a certain value each week, while other constraints are local, e.g., each member cannot receive more than certain number of emails each week.

## 3.2 Prediction with LTV

LinkedIn faces distinct challenges in predicting marketing performance due to the different nature of 2C and 2B segments, particularly the extended decision-making processes and delayed feedback for 2B products. To overcome this, we create separate models for predicting short-term metrics and LTV, then integrate the results to estimate the long-term metric. Here, the long-term metric $q^t$ is a numeric measure defined by marketers reflecting marketing performance over an extended period $t$, with short-term metrics representing observable metrics within a brief timeframe, like conversion and unsubscription. Notably, conversion is typically defined differently for 2B and 2C products. In 2B, it often refers to the submission of a qualified lead, whereas in 2C, a conversion can be marked by a customer signing up for a free trial.

To predict the short-term metrics, we develop predictive models forming a functional mapping:

$$y^{\text{conv}}_{u,a} = f_{\text{conv}}(x_u, x_a, x_c),$$
$$y^{\text{unsub}}_{u,a} = f_{\text{unsub}}(x_u, x_a, x_c),$$

where $y^{\text{conv}}$ and $y^{\text{unsub}}$ represent the predicted conversion probabilities and unsubscription probabilities respectively, $x_u$ denotes the feature vectors of users capturing user profiles, demographics, and other behavioral data, $x_a$ symbolizes the marketing actions undertaken, including the decision to send specific marketing emails to certain members, and $x_c$ represents other relevant contextual information that might influence the metrics.

We have a dedicated model to predict the LTV associated with a given conversion event. We calculate the LTVs as

$$y^{\text{ltv}}_{u,p} = f_{\text{ltv}}(x_u, x_p, x_c)$$

where $p$ indexes the product that email campaign $a$ tries to promote. We model the real-valued $y^{\text{ltv}}$ using regression equipped with a gamma loss function, defined as

$$L = \sum_i \left[ \mu_i k + q_i^{12} \exp(-\mu_i) k \right]$$

where $k$ is shape parameter, $i$ indexes the observations in training data, $q_i^{12}$ is the observed 12-month long-term metric (the ground truth value) and $\mu_i = log(y^{\text{ltv}}_i)$. This is a long time window that makes recent positive engagement unavailable for training (i.e., censored observations). In order to overcome this issue, we make auxiliary LTV predictions at shorter time windows (e.g., 1,3,6 months) and use the output of those models as features for the main 12-month LTV prediction. We found that this strategy increases prediction accuracy and allows the model to respond faster to evolving business trends.

## 3.3 Constrained and Multi-Objective Optimization

### 3.3.1 Formulating the Constrained Optimization Problem.
In our legacy system, a ranking heuristic was used in the email optimization stage. For each member, all candidate emails were ranked by a linear combination of $y^{\text{conv}}$ and $y^{\text{unsub}}$ with fixed coefficient $\kappa$, i.e., $y^{\text{conv}} - \kappa y^{\text{unsub}}$, and top-ranked emails were sent to the member with a fixed frequency cap. This approach did not optimize the business goal directly, and business constraints were only considered implicitly.

Here we apply the constrained optimization formulation in Section 2.2.2 to this problem. Let $u = 1, \ldots, U$ and $j = 1, \ldots, J$ index the member and campaigns respectively, and let $a_{u,j} \in \{0, 1\}$ denote the binary action decision of whether send email $j$ to member $u$. We can estimate the total expected long-term metric by $\sum_{u,j} a_{u,j} y_{u,j}^{\text{conv}} y_{u,p(j)}^{\text{ltv}}$, and total expected unsubscription by $\sum_{u,j} a_{u,j} y_{u,j}^{\text{unsub}}$.

Assuming that (i) we allow a maximum of $C_{\text{fcap}}$ emails to be sent to each member, (ii) we must send at least $C_{2B}$ and $C_{2C}$ emails for 2B and 2C products respectively, and (iii) we must have no more than $C_{\text{unsub}}$ unsubscription, we can form the constrained optimization problem as follows:

$$\min_{a_{u,j}} \sum_{u,j} -y_{u,j}^{\text{conv}} y_{u,p(j)}^{\text{ltv}} a_{u,j},$$

$$\text{s.t.} \sum_{u,j} a_{u,j} y_{u,j}^{\text{unsub}} \leq C_{\text{unsub}},$$

$$\sum_u \sum_{j \in \mathbb{J}_{2B}} -a_{u,j} \leq -C_{2B},$$

$$\sum_u \sum_{j \in \mathbb{J}_{2C}} -a_{u,j} \leq -C_{2C},$$

$$\sum_j a_{u,j} \leq C_{\text{fcap}}, \quad \forall u,$$

$$a_{u,j} \in \{0, 1\}, \quad \forall u, j.$$

where, $\mathbb{J}_{2B}$ and $\mathbb{J}_{2C}$ denote the email campaigns for 2B and 2C products respectively.

We recognize the optimization problem as an integer linear program (ILP), which can be combinatorially difficult to solve [36]. Instead, we relax the constraints on the $a_{u,j}$'s to allow them to take on values in the interval $[0, 1]$. The resulting problem is a linear program (LP) which is tractable (details in Section 3.3.2). We can interpret non-integer $a_{u,j}$ as the probability with which we should send an email about campaign $j$ to member $u$.

### 3.3.2 Solving the Large-Scale LP.
While there are open-source and commercial solvers for solving LPs, the size of the LP for our application is too large for these solvers. Recall that the optimization variable $a$ has dimension $UJ$, the product of the number of members and the number of campaigns. For a large social network like LinkedIn, we could have 10-100 millions of members and hundreds of email campaigns, resulting in tens of billions of variables. To solve the LP, we turn to *DuaLip* [4, 33, 34], a large-scale LP solver that we developed at LinkedIn. We briefly outline the algorithm here.

The LP in 3.3.1 can be written in the following form

$$\min_a y^\top a \quad \text{s.t.} \quad Da \leq b, \quad a_u \in \mathcal{B}_u, u \in \{1, \ldots, U\}, \quad (3)$$

where $a = (a_{1,1}, \cdots, a_{1,J}, \cdots, a_{U,J}) \in \mathbb{R}^{UJ}$ is the vector of decision variables, $y$ is the coefficient vector for the decision variables in the objective, $D$ is the design matrix in the constraints, $b$ is the constraint budget vector, and $\mathcal{B}_u$ are additional "simple" constraints, i.e., it is efficient to compute the Euclidean projection onto $\mathcal{B}_u$. In particular, $b = (C_{\text{unsub}}, -C_{2B}, -C_{2C})$, and $Da$ corresponds to the left-hand side of the first three constraints from the previous section. $\mathcal{B}_u$ corresponds to the remaining two constraints, i.e., box-cut constraints $\mathcal{B}_u = \{a_{u,j} : 0 \leq a_{u,j} \leq 1 \, \forall j, \, \sum_j a_{u,j} \leq C_{\text{fcap}}\}$.

Instead of solving the LP (3) directly, we solves a perturbed version of the LP, which is a quadratic program (QP):

$$\min_a y^\top a + \frac{\gamma}{2} a^\top a \quad \text{s.t.} \quad Da \leq b, \quad a_u \in \mathcal{B}_u, u \in \{1, \ldots, U\}, \quad (4)$$

where $\gamma > 0$ is a regularization hyperparameter. Dualizing just the polyhedral constraint $Da \leq b$ leads to the partial Lagrangian dual

$$g_\gamma(\lambda) = \min_{a \in \mathcal{B}} \left\{ y^\top a + \frac{\gamma}{2} a^\top a + \lambda^\top (Da - b) \right\},$$

where $\mathcal{B} = \prod_{u=1}^U \mathcal{B}_u$. Then the problem is transformed to calculate $\max_\lambda g_\gamma(\lambda)$, assuming strong duality holds. Applying Danskin's theorem [6], we know that $g_\gamma$ is differentiable and can explicitly compute its derivative:

$$\nabla g_\gamma(\lambda) = D\hat{a}(\lambda) - b,$$
$$\hat{a}_u(\lambda) = \Pi_{\mathcal{B}_u} \left[ -\frac{1}{\gamma}(D^\top \lambda + y)_u \right], \quad (5)$$

where $\Pi_{\mathcal{B}_u}(\cdot)$ is the Euclidean projection operator onto $\mathcal{B}_u$, and for a vector $z$ with the same dimensions as decision variable $a$, $z_u$ denotes the sub-vector with indices corresponding to member $u$. Notice that the computation of $\hat{a}(\lambda)$ can be parallelized across members $u$ and the box-cut projection $\Pi_{\mathcal{B}_u}$ can be computed efficiently by leveraging Wolfe's algorithm[40]. As a result, $\nabla g_\gamma(\lambda)$ can be computed efficiently. This allows us to use first-order methods such as accelerated gradient ascent [5] and L-BFGS [23] to obtain $\lambda^*_\gamma$, the dual value that optimizes the dual function $g_\gamma$. From this optimal dual value, we can obtain optimal primal values $a^*_\gamma$ via (5):

$$(a^*_\gamma)_u = \hat{a}_u(\lambda^*) = \Pi_{\mathcal{B}_u} \left[ -\frac{1}{\gamma}(D^\top \lambda^* + y)_u \right], \quad u = 1, \ldots, U.$$

While we obtain the optimal solution for the perturbed problem and not the original one, [4] shows that for sufficiently small $\gamma$ values, the perturbed solution is also optimal for the original problem:

*3.3.3 Practical Tips for Implementation.* We describe a couple of tips that make DuaLip converge much faster. (A) The choice of $\gamma$ is associated with a critical trade-off: DuaLip's solving of (4) is faster if $\gamma$ is larger, whereas DuaLip's approximate solution of (4) in a given number of iterations is closer to the solution of (3) if $\gamma$ is smaller. For one run, we try a grid of $\gamma$ values and choose the largest $\gamma$ value such that the regularizer part of the objective is small enough, i.e., the ratio, $\frac{\gamma a^\top a}{2|y^\top a|}$ is smaller than some threshold, say $10^{-3}$. After finding such a $\gamma$ value, we fix it and use it for all future production runs. (B) If $D$ is ill-conditioned (which typically happens when different constraints are on different scales), we can re-scale the constraints, i.e., pre-multiply $D$ and $b$ by a suitably chosen diagonal matrix.

With Dualip, the primal solution will fall on a vertex of $\mathcal{B}_u$ with high probability [34] when $\gamma$ is sufficiently small. This means majority of solutions are already binary, i.e., in $\{0, 1\}$. However, we still need to handle the cases when they are between 0 and 1. Simple rounding may violate the hard constraint $C_{\text{fcap}}$. We devise a randomization algorithm, basically sampling the campaign indexes with probability calculated from the primal solution (Appendix A.3). The randomization also provide additional benefit of balancing exploration-exploitation.

## 3.4 Audience Expansion

In the email marketing lifecycle, dedicated marketing teams from each business unit provide a list of candidate members for each email campaign. Historically, we have found these candidate member lists conservative in scope. To address this, we augment the provided lists with additional members through an Audience Expansion (AE) module in which we retrieve additional members who resemble the members provided by the marketing teams and append them to the candidate lists. We find such members via an approximate nearest neighbors (ANN) algorithm.

Formally, let $\mathcal{U}$ denote the overall member base at LinkedIn, let $\mathcal{J}$ denote the set of marketing campaigns, and let $\mathcal{U}_j^o \subset \mathcal{U}$ denote the set of original candidate members provided by a marketing team for email marketing campaign $j \in \mathcal{J}$. We leverage ANN via locality-sensitive hashing to produce a set

$$\mathcal{U}_j^{\text{expansion}} \subseteq \left\{u \in \mathcal{U}_j^- : \exists u' \in \mathcal{U}_j^o, \quad \text{s.t. } \|x_u - x_{u'}\|_2 \leq \delta\right\},$$

where, $\mathcal{U}_j^- = \mathcal{U} \setminus \mathcal{U}_j^o$ and $\delta$ is a hyperparameter which controls the size of the expansion audience. Ultimately, we combine the original audience with expansion audience for each campaign and input the expanded audience $\mathcal{U}^e$ into our optimization engine, where $\mathcal{U}^e = \bigcup_{j \in \mathcal{J}} \mathcal{U}_j^o \cup \mathcal{U}_j^{\text{expansion}}$.

We have found that as we relax $\delta$, there is a trade-off between wider campaign reach through increased expansion audience size and diminished incremental expansion audience quality. If $\delta$ is too large, a portion of the incremental expansion members added through AE are so dissimilar to the campaign's original members that they are never chosen by the LP to receive the marketing campaign. Moreover, the computational runtime of both Audience Expansion and the LP optimization increase in $\delta$, so choosing $\delta$ too large can lead to additional runtime for minimal gain. This latter concern about computational efficiency is especially important at LinkedIn's scale. On the other hand, if $\delta$ is too small then the expansion audience itself will be too small to produce a significant gain in downstream metrics of interest. We tune $\delta$ through offline evaluation to balance these trade-offs (Appendix D.3).

## 4 EXPERIMENT AND RESULT

### 4.1 Offline Analysis

We performed extensive offline analysis of prediction models, LP and AE. Due to space limitation, we only highlight the summary of offline analysis here and share the details of the results in Appendix D.

- **Prediction models evaluation.** We discuss classification and calibration performance of the two propensity models $f_{\text{conv}}$ and $f_{\text{unsub}}$ (Appendix D.1.1). We also evaluated the LTV model $f_{\text{ltv}}$ and find: (i) Gamma loss with auxiliary tasks can respond faster to market changes, (ii) it performs better than an alternative modeling approach using survival loss. (Appendix D.1.2)
- **Analysis on LP.** We find that smaller $\gamma$ value leads to better optimality but can take longer to converge (Appendix D.2.1). We also validated the high probability of binary primal solutions (Appendix D.2.2). More importantly, we investigated the fallback strategy of using previous dual solution for current LP, and find that we can maintain 99% value of the objective if data distribution is stable (Appendix D.2.3).
- **Analysis on AE.** We evaluated impact of $\delta$ and discuss the trade-off between audience size, quality and runtime (Appendix D.3).
- **Simulation before A/B test.** We find that (i) NOAH can make efficient "not-send" decisions through LP where legacy ranking system can not, and (ii) the combination of AE and LP helps realize the full potential of the optimization engine. (Appendix D.4)

### 4.2 Online Test Design

We design an online A/B test to compare the performance of the NOAH email system with that of the legacy system. In the test, we randomly assigned members to treatment arm (NOAH system) or control arm (legacy system). In addition, we included a third experimentation arm for the legacy system equipped with AE to isolate the impact of increasing marketing email volume on its own. The flow of the test is visualized in Figure 3.

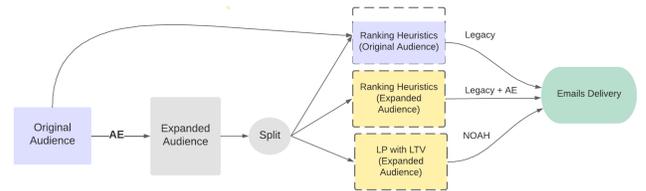

Figure 3: AB Test Flow

The email marketing only covers a small percentage of the member base. If the randomization splits on set of all members, the treatment effect could be diluted. We therefore focus on members that are "available" to receive marketing emails. In particular, we first randomize all members to 3 buckets, $\mathcal{U}_b^{\text{split}}$, where $b \in \{1, 2, 3\}$. For each week $t$ of the experiment, we can expand the original audience $\mathcal{U}_t^o$ to $\mathcal{U}_t^e$ with AE. Assuming the experiment runs for $T$ weeks, the set of members for experiment arm $b$ can be obtained as $\mathcal{U}_{b,T} = \bigcup_{t \leq T} \mathcal{U}_b^{\text{split}} \cap \mathcal{U}_t^e$. Note here that for control arm $b = 1$, the legacy system will rank and send emails only for members in original audience, i.e., over set of $\mathcal{U}_t^o \cap \mathcal{U}_1^{\text{split}}$ for week $t$.

Since it is not realistic to wait for 1 year long-term metric $q^{12}$, we define a long-term surrogate index [3] $\mathcal{Y}_u = \sum_{p \in \mathcal{P}_u} y_{u,p}^{\text{LTV}}$ for each member $u$, where $\mathcal{P}_u$ denote the set of products that member $u$ converted during the test period, and $y_{u,p}^{\text{LTV}}$ denotes the expected 1-year long-term metric associated with the conversion on product $p$ for member $u$. It is worth noting that although each observation consists of potentially multiple conversions, through offline analysis we have discovered that the majority of members convert at most one time during the test period.

Besides the above primary metric, we also monitor a few secondary metrics, e.g., email volume, conversions (Conv) as well as observed long-term metric. In addition, we have other metrics corresponding to the LP constraints, the key one being unsubscriptions (Unsub), a binary metric indicating whether or not a user unsubscribed during the test period.

We compare the means of above metrics from different arms for statistical testing of treatment effects. It worth noting that long-term metrics exhibit characteristics of a zero-inflated heavy-tail distribution (Appendix C.1). We performed simulations on type I error and power of t-test against these distributions, and found that the standard t-test is actually robust (Appendix C.2). We believe this result is broadly useful, as alternatives do not suit the needs of such testing: (1) Winsorization removes the large data points that actually matter (if not more), (2) Non-parametric tests (e.g., Mann-Whitney U test) lack a corresponding treatment effects.

### 4.3 Online Test Results

*4.3.1 Overview.* The A/B test results in major wins, one of the largest in Linkedin algorithmic marketing thus far. An overview of the results is presented in Table 1. All the numbers in the tables represent **relative** difference over the control arm, i.e., ratio of the difference (between the treatment and control arms) to the mean value in control. Results that are not statistically significant are marked either "neutral" or "not stat sig".

Table 1: Overview of A/B Test Result

| Marketing System | Surrogate Index | Unsub | Email Volume |
| --- | --- | --- | --- |
| Legacy+AE | neutral | +8.08% | +9.42% |
| NOAH | +2.16%[1] | +1.52% | +8.44% |

[1] All percentages are relative to the control arm.

As shown in Table 1, NOAH produces a statistically significant positive lift in expected 1-year long-term metric (Surrogate Index) at the cost of a moderate uptick (+1.52%) in total unsubscription. Given that Email Volume is increased by 8.44%, the unsubscription per email sent is actually reduced, indicating that more relevant

Table 2: Product Type Breakdown

| Product Type | Surrogate Index | 3-Mth Metric[1] | Conv |
| --- | --- | --- | --- |
| 2C | not stat sig | neutral | neutral |
| 2B | +13.21% | +23.66% | +13.30% |

[1] Observed 3-month long-term metric $q^3$ after test concluded.

emails were sent. By comparison, the legacy ranking system with AE (Legacy+AE) does not produce a statistically significant effect on the long-term metric, and it produces large Unsub increment (+8.08%), a magnitude similar as Email Volumne increment (+9.42%).

*4.3.2 Product Type Breakdown.* Historically, we have faced challenges in developing a unified marketing system which optimizes marketing emails for a heterogeneous suite of 2C and 2B products. Our legacy marketing system tends to optimize for quick sign-up 2C products in spite of these products' relatively small lifetime values. On the other hand, campaigns associated with high-LTV 2B products tend to be neglected.

A breakdown of comparison between NOAH and control into 2C and 2B product types is presented in Table 2. We observe that NOAH is able to drive a large positive lift in both Surrogate Index (+13.21%) and total Conv (+13.30%) for high-LTV 2B products without sacrificing these metrics for 2C products. It achieves this lift by increasing total 2B conversions and shifting the distribution 2B conversions toward those with higher Surrogate Index, as visualized in Appendix Figure 14.

Due to the nature of delayed feedback, we tracked the long-term metric of conversions happened during A/B test after the test concluded. We find that NOAH is able to drive a significant lift (+23.66%) in observed 3-month long-term metric (3-Mth Metric) for 2B products. Similarly to Surrogate Index, we observed distributional shift toward high-value conversions in 3-Mth Metrics, as shown in Appendix Figure 15.

The breakdown analysis demonstrates that NOAH is able to lift high value long-term metrics for 2B products while simultaneously maintaining 2C acquisition metrics and controlling Unsub. This holistic behavior is essential for a unified marketing system which must support a wide range of campaigns having different downstream value.

## 5 CONCLUSION

In this paper, we introduce various challenges faced when building an intelligent marketing system and propose a general framework, NOAH, to address these challenges. We give guidelines on how to apply the NOAH framework and provided a few high-level examples. In addition, we report a large-scale application of NOAH to LinkedIn's email marketing use case. We discuss important innovations and learnings in the implementation and share results of the major wins for the corresponding online A/B test.

## ACKNOWLEDGMENTS

We thank Shruti Sharma, Maggie Zhang, Divyakumar Menghani, Aarthi Jayaram, and Rehan Khan for their support of the work. We also thank our colleagues Michael Ingley, Grinishkumar Engineer,

Artem Baranov, Sumukh Sagar Manjunath, Carlos Flores, Shawn Bianchi, Kim Foo, Kyla Falkenhagen for their collaboration on the email application and A/B test. Additionally, we'd like to acknowledge Zhiqi Guo, Ming Wu, Miao Cheng and Faisal Farooq for their contributions and support during the early phase of the work.

# A ALGORITHMS

## A.1 Bidding Optimization with Lagrangian Relaxation

We introduce the constraint into the objective with Lagrangian relaxation:

$$g(\lambda) = \max_{p(u) \in \{1,...,P\}} \sum_u y^{\text{rev}}_{u,p(u)} - \lambda \left( C_{ROAS} \sum_u y^{\text{cost}}_{u,p(u)} - \sum_u y^{\text{rev}}_{u,p(u)} \right)$$

$$= \sum_u \max_{p(u) \in \{1,...,P\}} \left[ (1+\lambda) y^{\text{rev}}_{u,p} - \lambda C_{ROAS} y^{\text{cost}}_{u,p} \right]$$

For any fixed $\lambda = \lambda'$, we can get the optimal $p(u)$ simply by finding the maximum among $P$ discrete values for each $u$, i.e.,

$$p_{\lambda'}(u) = \underset{p}{\arg\max} \{ (1+\lambda') y^{\text{rev}}_{u,p} - \lambda' C_{ROAS} y^{\text{cost}}_{u,p} \}, \quad \forall u.$$

Then, we can compute the subgradient with respect to $\lambda$,

$$\partial g(\lambda)|_{\lambda=\lambda'} = \sum_u y^{\text{rev}}_{u,p_{\lambda'}(u)} - C_{ROAS} \sum_u y^{\text{cost}}_{u,p_{\lambda'}(u)}$$

and use the sign of the subgradient to guide where to find $\lambda$ that minimize $g(\lambda)$.

---

**Algorithm 1** Bidding Optimization with Lagrangian Relaxation

**Input:** $y^{\text{rev}}_{u,p}$, $y^{\text{cost}}_{u,p}$, $C_{ROAS}$, $\lambda_{\text{init}} > 0$, and $0 < \epsilon \ll \lambda_{\text{init}}$
**Output:** $p^*(u)$ for each $u$
  set $\lambda_{min} \leftarrow 0$ and $\lambda_{max} \leftarrow \lambda_{\text{init}}$
  **if** $\partial g(\lambda)|_{\lambda=0} \geq 0$ **then**
    return $p_0(u)$
  **else**
    **while** $\partial g(\lambda)|_{\lambda=\lambda_{max}} < 0$ **do**
      set $\lambda_{min} \leftarrow \lambda_{max}$ and $\lambda_{max} \leftarrow 2\lambda_{max}$
    **end while**
  **end if**
  **while** $\lambda_{max} - \lambda_{min} > \epsilon$ **do**
    set $\lambda' \leftarrow \frac{\lambda_{min}+\lambda_{max}}{2}$
    **if** $\partial g(\lambda)|_{\lambda=\lambda'} > 0$ **then**
      set $\lambda_{max} \leftarrow \lambda'$
    **else if** $\partial g(\lambda)|_{\lambda=\lambda'} < 0$ **then**
      set $\lambda_{min} \leftarrow \lambda'$
    **else**
      break
    **end if**
  **end while**
  return $p_{\lambda'}(u)$

---

## A.2 Content Optimization with Online Contextual Bandit

For efficient Bayesian online learning, we will assume the priors for all weights are independent and Gaussian-distributed. The joint distribution can be written in the following form:

$$p(y, t, s, w_1, w_2, \cdots, w_k) = h(y,t)a(t,s)g(w_1, w_2, \cdots, w_k, s) \prod_i r_i(w_i),$$

where

$$r_i(w_i) = \mathcal{N}(w_i; \mu_i, \sigma_i^2),$$
$$g(w_1, \cdots, w_K, s) = \delta(x^T w = s),$$
$$a(s, t) = \mathcal{N}(t; s, \beta^2),$$
$$h(y, t) = \delta(\rho(t) = y),$$

and $\delta(\cdot)$ is the indicator function.

In the factor graph, the variable nodes $\{w_i, s, t, y\}$ are connected by the factor nodes $\{r_i, g, a, h\}$. We can calculate the posterior distribution $p(w_i|x, y)$ by message passing in factor graph:

(1) Consider $t$ as root node.
(2) Perform message passing from leaf to root and obtain $p(t) = \mathcal{N}(t; \eta, \tau^2) h(y, t)$, where $\eta = \sum_i x_i \mu_i$, and $\tau^2 = \sum_i x_i^2 \sigma_i^2 + \beta^2$.
(3) If $p(t)$ is not Gaussian, we use moment matching to obtain $\hat{p}(t) = \mathcal{N}(t; \mu_t, \sigma_t^2)$, where $\mu_t = \eta + \Delta$ and $\sigma_t^2 = \tau^2 \gamma$.
(4) Perform message passing from root to leaf and obtain posterior of model weights,

$$p(w_i) = \mathcal{N}(w_i; \mu_{w_i}, \sigma_{w_i}^2),$$

where,

$$\begin{cases} \sigma_{w_i}^2 &= \sigma_i^2 [1 - x_i^2 \frac{\sigma_i^2}{\tau^2}(1-\gamma)] \\ \mu_{w_i} &= \mu_i + x_i \frac{\sigma_i^2}{\tau^2} \Delta \end{cases}$$

With the above, we can have following online contextual bandit algorithm that works for high-dimensional settings. The algorithm can be easily parallelized or set up for on-device computing.

---

**Algorithm 2** Content Optimization with Online Contextual Bandit

**Input:** $\mu_i, \sigma_i^2, \beta^2$
  **for** time step $j = 1, 2, \cdots, J$ **do**
    Sample $\hat{w}_i$ from $N(\mu_{w_i}, \sigma_{w_i}^2)$ for all $i$;
    **for** creative $a = 1, 2, \cdots, A$ **do**
      Sample $\hat{\epsilon}_a$ from $N(0, \beta^2)$
      Calculate latent reward $\tilde{t}_{j,a} = \hat{\epsilon}_a + \sum_i x_{i,j,a} \hat{w}_i$
    **end for**
    Choose creative $a_j = \arg\max_a \tilde{t}_{j,a}$ and show it to user;
    observe user engagement and receive reward $y_j$
    Update $\mu_{w_i} \leftarrow \mu_{w_i} + x_i \frac{\sigma_{w_i}^2}{\tau^2} \Delta$;
    Update $\sigma_{w_i}^2 \leftarrow \sigma_{w_i}^2 [1 - x_i^2 \frac{\sigma_{w_i}^2}{\tau^2}(1-\gamma)]$
  **end for**

---

Here, the value of $\gamma$ and $\Delta$ depends on the link function. For linear regression, $\gamma = 0$ and $\Delta = y - x^T w$. For Probit regression, $\gamma = 1 - \omega(\frac{y\eta}{\tau})$ and $\Delta = y\tau\vartheta(\frac{y\eta}{\tau})$, where, $\vartheta(t) = \frac{\mathcal{N}(t;0,1)}{\Phi(t;0,1)}$ and $\omega(t) = \vartheta(t)(\vartheta(t) + t)$.

## A.3 Sampling from Primal Solution

The IP to LP relaxation can be interpreted as transforming the deterministic integer decision to the expected value of a random integer decision, i.e., the probability of sending some marketing email. So naturally we can add a randomization after the LP solution, to "recover" the random integer decision $a'_{u,j}$. Formally, we want to implement the sampling so that:

$$E(a'_{u,j}) \propto a_{u,j}$$
$$\sum_j a'_{u,j} \leq C_{\text{fcap}}$$

We can follow the following algorithm to "recover" the random integer decision $a'_{u,j}$:

(1) Calculate sampling probabilities $p_{u,j} = \frac{a_{u,j}}{\sum_j a_{u,j}}$
(2) Sample a set $\mathcal{J}'_u$ of size $\lfloor \sum_j a_{u,j} + 0.5 \rfloor$ without replacement, from set of all index $\mathcal{J}_u$ with non-uniform distribution of $p_{u,j}$
(3) We obtain $a'_{u,j} = I(j \in \mathcal{J}'_u)$, where $I(\cdot)$ is indicator function.

## B ADAPTIVE HEURISTICS

There are two main classes of AH: 1) Feedback Loop Controller and 2) Hierarchical Optimization. At a high-level, these two classes of AH help optimize and stabilize the system on the "time" and "space" dimensions respectively.

Besides the two main classes of AHs, there are other simple AHs, for example: cool-off rules so that a member cannot receive more than some number marketing messages over a fixed time window, and $\epsilon$-greedy exploration so that selection bias is measured and taken into account in modeling. Most of these heuristics are common to RecSys, and we refer readers to standard practice in [8, 10].

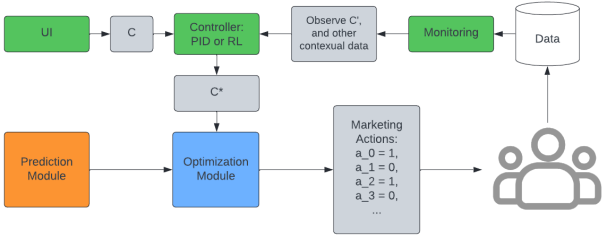

Figure 4: Feedback Loop Controller

### B.1 Feedback Loop Controller

Controllers can observe the feedback from the marketing units, and is able to adjust input to the main optimization module, so as to stabilize it when there are deviations on the constraints. This is particularly useful when we have strict constraints, e.g., marketing budgets/cost. Technically, this can be achieved either by a Proportional–integral–derivative (PID) controller and/or a lightweight reinforcement learning (RL) model. Figure 4 shows the overall flow.

Without loss of generality, assume there is a marketing cost metric $C_1$ we want to monitor and control from the constraint formulation (2) in Section 2.2.2. The controller takes observed cost $C'_1$, and other historical data $x^H$, and outputs the cost target $C^*_1$ which should be used as the budget in the linear program, i.e.,

$$C^*_1 = g(C_1, C'_1, \phi(x^H)).$$

For the case of a PID controller, $\phi(x^H)$ represents the three quantities, proportion, integral, and derivative, which are derived from the time series of error term $e_t(x^H) = C'_{1,t} - C_{1,t}$. Once $C^*_1$ is obtained, we can use it for main optimization module:

$$\max_{a_u} \sum_u f_0(x_u, x_{a_u}, x_c),$$
$$\text{s.t.} \sum_u f_1(x_u, x_{a_u}, x_c) \leq C^*_1,$$
$$\sum_u f_k(x_u, x_{a_u}, x_c) \leq C_k, \quad \forall 2 \leq k \leq K,$$

Beside PID controllers, we can also use RL to learn more complex $g(.)$ and $\phi(.)$ functions.

### B.2 Hierarchical Optimization

The formulation of Hierarchical Optimization is similar to the LP formulations in Section 2.2.2, except that 1) we use aggregated data as input, 2) the prediction model can be light-weight, and 3) the optimization problem is small-scale.

The main use case for hierarchical optimization is macro-level marketing budget allocation. With slight abuse of notation from Section 2.2.2, we let $u$ index unit of budget allocation (e.g., combination of marketing channel and lines of business), and let $a_u$ denote the level of budget to be assigned to unit $u$. The resulting optimization formulation is the same as (1) and (2).

The model used in the objective is typically the revenue on budget changes $y^0 = f_0(x_u, x_{a_u}, x_c)$, where either traditional Media Mixed Models (MMM) or modern causal ML methods can be used. Among the constraints, the cost model $y^1 = f_1(x_u, x_{a_u}, x_c)$ can either use learned or rule-based $f_1(.)$ (e.g., assuming cost equals fixed proportion of budget) depending on the maturity of the optimization engine. The constrained optimization problem is small-scale, so we can typically use off-the-shelf solvers to solve the original integer program without LP relaxation.

## C ROBUSTNESS OF T-TEST

### C.1 Zero-Inflated Heavy-Tail Metric

We can see that in Figure 5, there is a significant probability mass at zero and right side of the distribution is heavy-tailed for Surrogate Index. Note that the counts on histogram plot is $log_{10}$ scale.

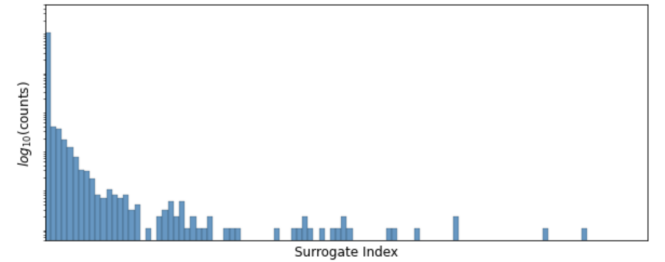

Figure 5: Distribution of Surrogate Index

### C.2 Simulation Result for t Test Robustness

We simulate the control and treatment based on zero-inflated heavy-tail data. The control is simulated by $X = Z_x X'$, where $Z_x$ is random binary variable with $\mathbb{E}[Z_x] = p_0$, and $X'$ is sampled from a heavy-tail distribution. The treatment is simulated by $Y = Z_y Y'$, where $\mathbb{E}[Z_y] = p_0 + p_\Delta$ and $\mathbb{E}[\log(Y')] = \mathbb{E}[\log(X')] + \Delta$.

The null model is simulated by setting $\Delta = 0$ and $p_\Delta = 0$. Besides varying the size of the test ($\alpha$), we also (i) vary $p_0$ at 0.02 and at 0.8, (ii) vary distribution of $X'$ at $P_{\text{obs}}$ and $P_{\text{ln}}$. Here $P_{\text{obs}}$ is empirical distribution of positive Surrogate Index in AB test, and $P_{\text{ln}}$ is log normal distribution with $\mu(\log(X')) = 2$ and $\sigma(\log(X')) = 2$. We perform t-test on the simulated samples and calculated the estimated type 1 errors. We can see type 1 errors are well controlled as shown in Table 3.

**Table 3: Type 1 errors of t test against zero-inflated heavy-tail distribution**

| $\alpha^1$ | $X' \sim P_{\text{obs}}$ | | $X' \sim P_{\text{ln}}$ | |
|---|---|---|---|---|
| | $p_0 = 0.02$ | $p_0 = 0.8$ | $p_0 = 0.02$ | $p_0 = 0.8$ |
| 0.05 | 0.03038 | 0.04837 | 0.02464 | 0.03917 |
| 0.01 | 0.00227 | 0.00858 | 0.00168 | 0.00524 |
| 0.005 | 0.00072 | 0.00376 | 0.00048 | 0.00211 |
| 0.001 | 0.00002 | 0.0005 | 0.00002 | 0.00031 |

[1] Size of the test.

The alternative model is simulated by varying value of $\Delta$ and $p_\Delta$. For simplicity, we fix $\alpha = 0.05$, $p_0 = 0.2$ and vary distribution of $X'$ at $P_{\text{obs}}$ and $P_{\text{ln}}$. We can see in Table 4 that power increases as $p_\Delta$ increases or as $\Delta$ increases.

**Table 4: Power of t test against zero-inflated heavy-tail distribution**

| $p_\Delta$ | $\Delta$ | $X' \sim P_{\text{obs}}$ | $X' \sim P_{\text{ln}}$ |
|---|---|---|---|
| 0 | 1.0 | 0.451 | 0.153 |
| 0 | 2.0 | 0.909 | 0.496 |
| 0.005 | 1.0 | 0.673 | 0.295 |
| 0.005 | 2.0 | 0.976 | 0.637 |

## D OFFLINE ANALYSIS FOR EMAIL APPLICATION

### D.1 Prediction Models Evaluation

*D.1.1 Propensity Model Evaluation.* We evaluate our propensity models $y_{u,a}^{\text{conv}} = f_{\text{conv}}(x_u, x_a, x_c)$ and $y_{u,a}^{\text{unsub}} = f_{\text{unsub}}(x_u, x_a, x_c)$ using standard offline metrics for binary classification models, such as area under the ROC curve (AUC) and area under the precision-recall curve (AUPR). Average values for these metrics are provided in Table 5. ROC curves for representative training runs of $f_{\text{conv}}$ and $f_{\text{unsub}}$, respectively, are shown in Figure 7. These models are re-trained on a regular cadence to account for distributional shifts resulting from evolving member engagement patterns and updated marketing campaigns.

**Table 5: Propensity model offline performance**

| Propensity Model | AUC | AUPR |
|---|---|---|
| $f_{\text{conv}}$ | 0.895 | 0.002 |
| $f_{\text{unsub}}$ | 0.715 | 0.002 |

We also evaluate the calibration performance of our propensity models, as their output estimates serve as the inputs to the LP. Miscalibration in these estimates can distort the LP's ability to optimize accurately, leading to sub-optimal downstream performance. In practice, we find varying calibration performance from our propensity models. To address this, we can include an additional calibration layer via isotonic regression to improve calibration performance. We can then monitor offline metrics such as Brier score and log-loss before and after calibration to validate the improvement. An example calibration plot for $f_{\text{unsub}}$ is shown in Figure 8.

*D.1.2 LTV Model Evaluation.* We evaluate the performance of LTV models by comparing the predictions to the observed long term metric $q^{12}$ of previous conversion.

To handle "censored observations" for recent conversions, we leverage auxiliary tasks modeling metric observed for 1 month, 3 month and 6 months, i.e., $q^1$, $q^3$ and $q^6$, and use the predicted value from auxiliary tasks as input to the main task. We found that model with auxiliary tasks has lower prediction error and more importantly, can respond faster to evolving trends on long-term metric as shown in Figure 9.

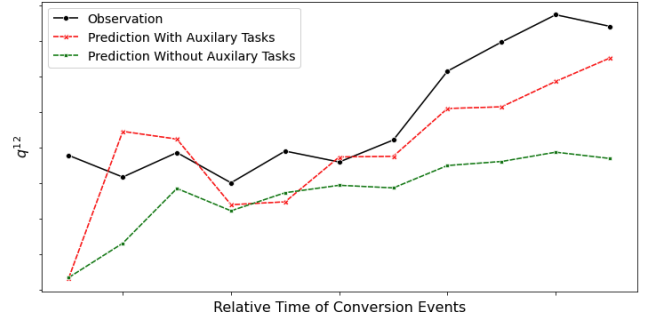

**Figure 9: Gamma model with and without auxiliary tasks**

Alternatively, we can also handle the censored observation by survival modeling. The long-term metric can be modeled as a product of survival time and the long-term metric generated per time unit, where the second term is modeled using mean squared error loss. The survival time can be modeled with accelerated failure rate loss, assuming latent survival time followed log-logistic distribution, i.e.,

$$l_i = \begin{cases} \log(2 + e^{\frac{s_i - \mu_i}{\sigma}} + e^{\frac{\mu_i - s_i}{\sigma}}) & \text{if } s_i \text{ is observed} \\ \frac{s_i - \mu_i}{\sigma} + \log(1 + e^{\frac{\mu_i - s_i}{\sigma}}) & \text{if } s_i \text{ is right censored} \end{cases}$$

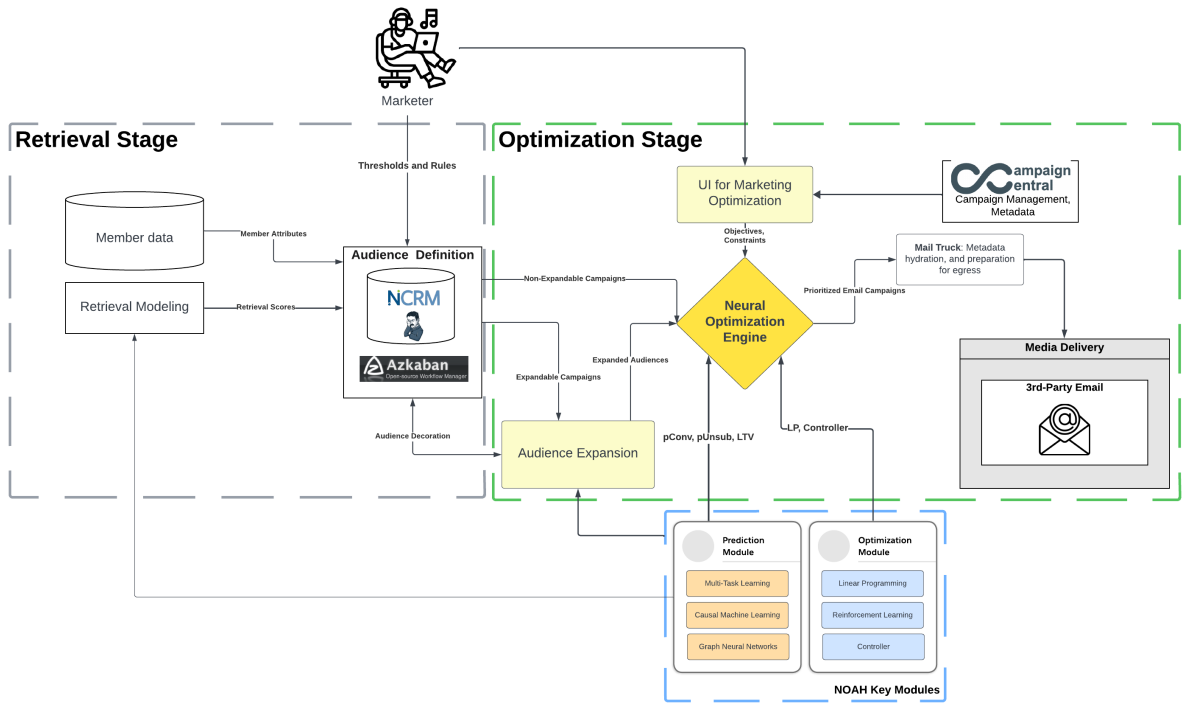

Figure 6: Two-stage Email Marketing System with NOAH

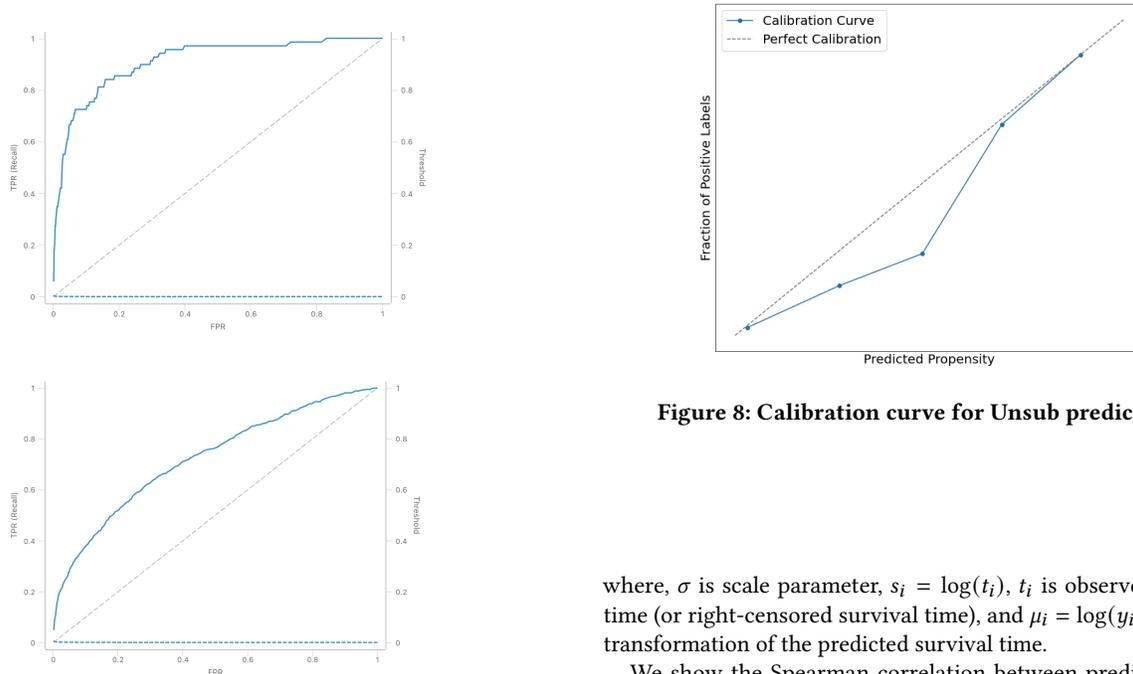

Figure 7: ROC curves for Conv and Unsub predictions

Figure 8: Calibration curve for Unsub predictions

where, $\sigma$ is scale parameter, $s_i = \log(t_i)$, $t_i$ is observed survival time (or right-censored survival time), and $\mu_i = \log(y_i)$ is the log transformation of the predicted survival time.

We show the Spearman correlation between predictions and $q^{12}$ for 5 representative product groups under the Survival model and the Gamma model (with auxiliary tasks) in Table 6. Overall, the Gamma model has better performance for most of the product groups. We also show the calibration curve of LTV predictions vs $q^{12}$ observation in Figure 10.

Table 6: Spearman rank correlation of Survival and Gamma Models

| Product Group | Survival | Gamma | Rel. Diff. |
| --- | --- | --- | --- |
| 2C #1 | 0.448 | 0.491 | +9.6% |
| 2C #2 | 0.526 | 0.557 | +5.9% |
| 2C #3 | 0.425 | 0.489 | +15% |
| 2B #1 | 0.336 | 0.369 | +9.8% |
| 2B #2 | 0.458 | 0.396 | -13% |

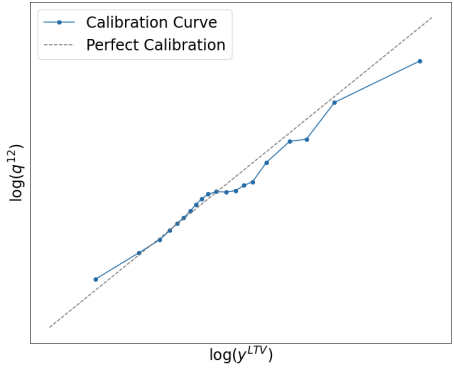

Figure 10: Calibration curve for LTV prediction

## D.2 Analysis on LP

*D.2.1 Impact of $\gamma$ values.* For the LP in Section 3.3.1, the two items that are important are: (a) the value of the primal objective that is being maximized; and (b) how well the constraints are satisfied. Given a constraint $d_i^T a \leq b_i$, we can define a (normalized) quantity, $v_i = \max\{0, (b_i - d_i^T a)/(1 + |b_i|)\}$ that says how much the $i$-th constraint is violated. With many constraints, we can define a single violation score called *feasibility* as $\max_i v_i$. Figure 11 gives the DuaLip trajectories *(primal objective, feasibility)* for various $\gamma$ values. The vertical line defines the maximum allowed feasibility violation for our application. Thus, even though during a large part of an optimization trajectory the primal objective has large values, they are unacceptable from the feasibility angle. We would choose the point having the largest primal objective among the acceptable feasible points to the left of the vertical feasibility line. As expected, smaller $\gamma$ values yield much better solutions.

*D.2.2 Vertex primal solutions.* The primal solution $a_{u,j}$ for a given member $u$ and given email campaign $j$ will be binary with high probability. We ran Dualip with $\gamma = 10^{-8}$ for multiple different datasets from different weeks. Averaging across these datasets, the percentage of members with a binary primal solution was 95.81% (S.E. 0.61%).

*D.2.3 Freshness of dual solution.* There is a service-level-agreement (SLA) that the optimization stage has to be completed within a certain time so that marketing emails can be sent out as planned. However, there are times when the LP algorithm takes longer than the SLA to converge. As a fallback strategy when this happens, we use the dual solution ($\lambda$) saved from the previous period to compute the primal solution for the current period using (5). The assumption is that optimality is not impacted much if the data distribution of the inputs of the LP does not vary much across weeks.

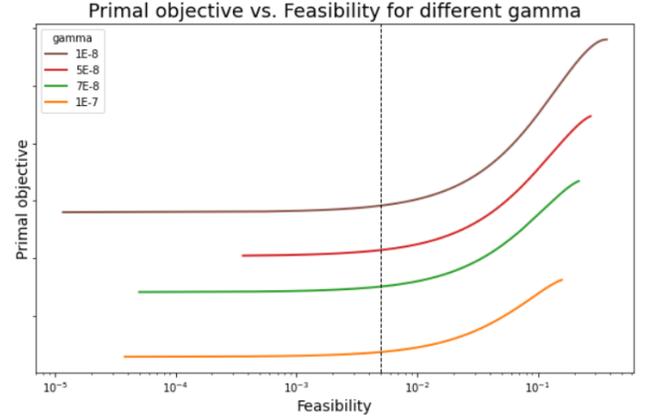

Figure 11: Primal objective vs. feasibility for different $\gamma$ values

We investigate this assumption by using datasets from consecutive weeks. We first validated the stableness of the data distribution week-over-week using the Kolmogorov-Smirnov (KS) statistic (Figure 13). We then ran the Dualip algorithm for each of the consecutive weeks. For each week, we compare the primal objective value, i.e., $\sum_{u,j} a_{u,j} y_{u,j}^{\text{conv}} y_{u,p(j)}^{\text{ltv}}$, achieved by the dual solution from Week 0 against that achieved by the current dual solution (Figure 12). The primal objective achieved by Week 0's dual never dropped below 99% of that achieved by the current week's dual.

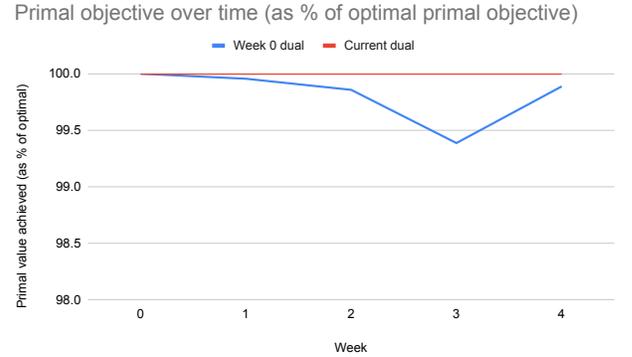

Figure 12: Comparison of primal objective achieved by current and Week 0 dual solution

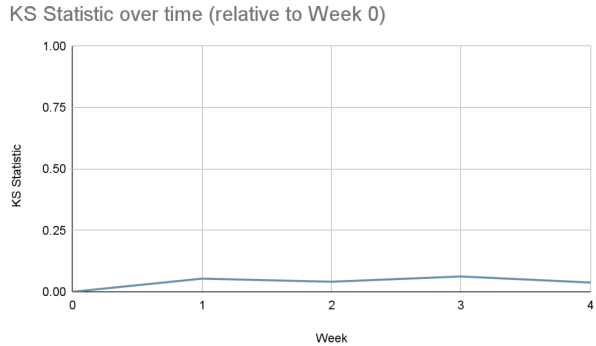

Figure 13: Kolmogorov-Smirnov (KS) statistic of data distribution from week $t$ relative to that of week 0

## D.3 Analysis on AE

The hyperparameter $\delta$ controls the size of the incremental expansion audience. As we increase $\delta$, both the size of the expansion audience and the overall runtime of the engine increase, but there is diminishing marginal audience quality. The relationship between $\delta$ and these variables is shown in Table 8.

Table 7: Influence of $\delta$ on expanded audience size, LP output size, and overall runtime

| $\delta$ | Audience Size | LP Output Size | Overall Runtime |
| --- | --- | --- | --- |
| 0.00001 | +0.38% | +0.33% | +4.21% |
| 0.00005 | +1.04% | +0.95% | +4.77% |
| 0.0001 | +3.77% | +3.23% | +15.42% |
| 0.0005 | +28.42% | +8.44% | +16.11% |
| 0.001 | >+250% | +9.59% | >+300% |

Audience size and overall runtime increase dramatically in $\delta$, but the LP output size plateaus. This indicates that the quality of the incremental expansion audience decreases significantly as we relax $\delta$.

## D.4 Simulation before A/B Test

Prior to the launch of the actual A/B test, we conduct offline evaluations of our optimization engine using historical data. This assessment aims to understand the engine's performance comprehensively and allow for necessary adjustments, particularly to the parameters of LP constraints.

*D.4.1 Four-Way Comparison.* The evaluation is designed to compare four distinct scenarios to understand the optimization engine's efficacy fully:

- Legacy ranking system with the original audience list (Legacy).
- Legacy ranking system with the AE list (Legacy with AE).
- NOAH optimization system with the original list (NOAH without AE).
- NOAH optimization system with the AE list (NOAH).

*D.4.2 Simulation Result.* The comparison results are shown in Table 8 and 9. All the numbers in the table are relative to the performance of the Legacy ranking system.

Table 8 shows the relative number of members that receives 0, 1 and 2+ Emails. It worth noting that the comparison is based on counterfactual audience from AE (details in 4.2). We can see that the rate of sending "0 Emails" is much higher for the two scenarios that utilized LP (+92% for NOAH without AE and +158% for NOAH). This indicates that LP optimization can make efficient "not-send" decision, so that it can reduce spamming and improve member experience. In fact, the ranking heuristics itself can't make "not-send" decision at all.

From Table 9, we can see that NOAH without AE can reduce Unsub rate by 9%, but only improved long-term metric by 2%. This suggests that LP's potential is limited by the original list's audience coverage. Meanwhile, with AE widens the top funnel, Legacy ranking system could not control Unsub rate (+9%). It is the combination of LP and AE that realize the optimization of long-term metrics (+7%) without negatively impacting member experience.

Table 8: Email distribution

| Scenario | 2+ Emails | 1 Email | 0 Emails |
| --- | --- | --- | --- |
| Legacy | 1.00 | 1.00 | 1.00 |
| Legacy with AE | 1.19[1] | 1.08 | 0.00 |
| NOAH without AE | 0.99 | 0.99 | 1.92 |
| NOAH | 1.16 | 1.05 | 2.58 |

[1] all values are relative to the Legacy system.

Table 9: Expected long-term metrics and Unsub rates

| Scenario | Emails Volume | Long-term Metric | Unsub. Rate |
| --- | --- | --- | --- |
| Legacy | 1.00 | 1.00 | 1.00 |
| Legacy with AE | 1.12[1] | 1.05 | 1.09 |
| NOAH without AE | 0.99 | 1.02 | 0.91 |
| NOAH | 1.09 | 1.07 | 0.92 |

[1] all values are relative to the Legacy system.

## E  ADDTIONAL A/B TEST RESULTS

### E.1  Distribution of Long-Term Metrics on 2B Conversions

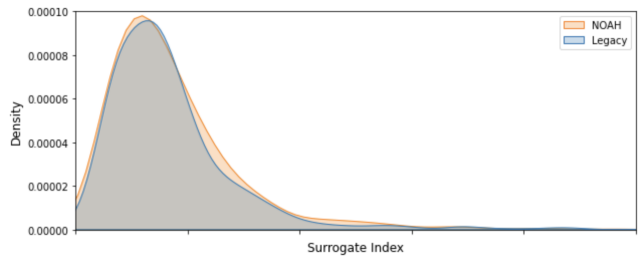

Figure 14: Distribution of Surrogate Index for 2B Conversions

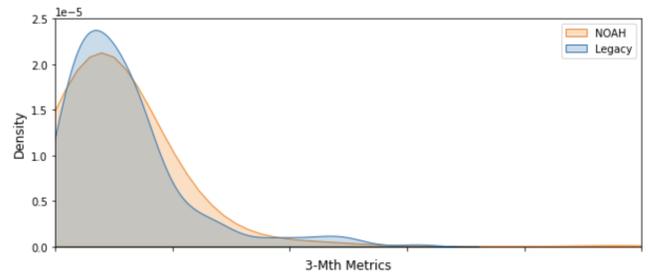

Figure 15: Distribution of 3-Month Metric for 2B conversions